\documentclass{Interspeech}

\usepackage{amsmath,amsfonts}
\usepackage{amssymb}
\usepackage{mathtools}
\usepackage{algorithm}
\usepackage{algorithmic}
\usepackage[linesnumbered,ruled, algo2e]{algorithm2e}
\usepackage{setspace, fullwidth}
\usepackage{lipsum}
\usepackage{array}
\usepackage[caption=false,font=normalsize,labelfont=sf,textfont=sf]{subfig}
\usepackage{textcomp}
\usepackage{stfloats}
\usepackage{url}
\usepackage{verbatim}
\usepackage{graphicx}
\usepackage{cite}
\usepackage{orcidlink}
\usepackage{multirow}
\usepackage{multicol}
\usepackage{booktabs}
\usepackage{tcolorbox}
\usepackage{flushend}
\usepackage{makecell}
\usepackage{colortbl}
\usepackage{color}
\usepackage{threeparttable} 
\usepackage{dsfont}
\usepackage{amsmath,amssymb}
\usepackage{relsize}
\usepackage{xcolor}

\interspeechcameraready

\title{Bona fide Cross Testing Reveals Weak Spot in Audio Deepfake Detection Systems}

\author[affiliation={1,2}]{Chin Yuen}{Kwok}
\author[affiliation={2}]{Jia Qi}{Yip}
\author[affiliation={1}]{Zhen}{Qiu}
\author[affiliation={1}]{Chi Hung}{Chi}
\author[affiliation={1}]{Kwok Yan}{Lam}

\affiliation{Digital Trust Centre}{Nanyang Technological University}{Singapore}
\affiliation{College of Computing and Data Science}{Nanyang Technological University}{Singapore}
\email{kwok0062@e.ntu.edu.sg}

\usepackage{comment}

\begin{document}

\maketitle

\begin{abstract}

Audio deepfake detection (ADD) models are commonly evaluated using datasets that combine multiple synthesizers, with performance reported as a single Equal Error Rate (EER). However, this approach disproportionately weights synthesizers with more samples, underrepresenting others and reducing the overall reliability of EER. Additionally, most ADD datasets lack diversity in bona fide speech, often featuring a single environment and speech style (e.g., clean read speech), limiting their ability to simulate real-world conditions. To address these challenges, we propose bona fide cross-testing, a novel evaluation framework that incorporates diverse bona fide datasets and aggregates EERs for more balanced assessments. Our approach improves robustness and interpretability compared to traditional evaluation methods. We benchmark over 150 synthesizers across nine bona fide speech types and release a new dataset to facilitate further research at \url{https://github.com/cyaaronk/audio_deepfake_eval}.

\end{abstract}

\section{Introduction}

Audio deepfake detection (ADD) focuses on identifying synthetic or manipulated audio, commonly referred to as spoofed audio, which aims to replicate genuine recordings. These deepfakes, created using advanced machine learning techniques, pose significant risks to systems relying on voice-based authentication, media forensics, and public trust, especially in the context of misinformation campaigns. The development of robust ADD models is thus critical to mitigate these threats.

Currently, ADD models are primarily evaluated on a single dataset that combines multiple synthesizers \cite{kwok2025robust,liu2023asvspoof,muller2024mlaad} or through spoof cross-testing \cite{wu2024codecfake,Wang2019ASVspoof2A,zhao2024emofake}, a method where bona fide audio is paired with $M$ spoof data subsets (one subset corresponds to one synthesizer) to generate $M$ test sets. For each test set, performance is typically reported in terms of Equal Error Rates (EERs). However, this evaluation approach suffers from two key limitations.

\begin{figure}[t]
  \centering
  \includegraphics[width=\linewidth]{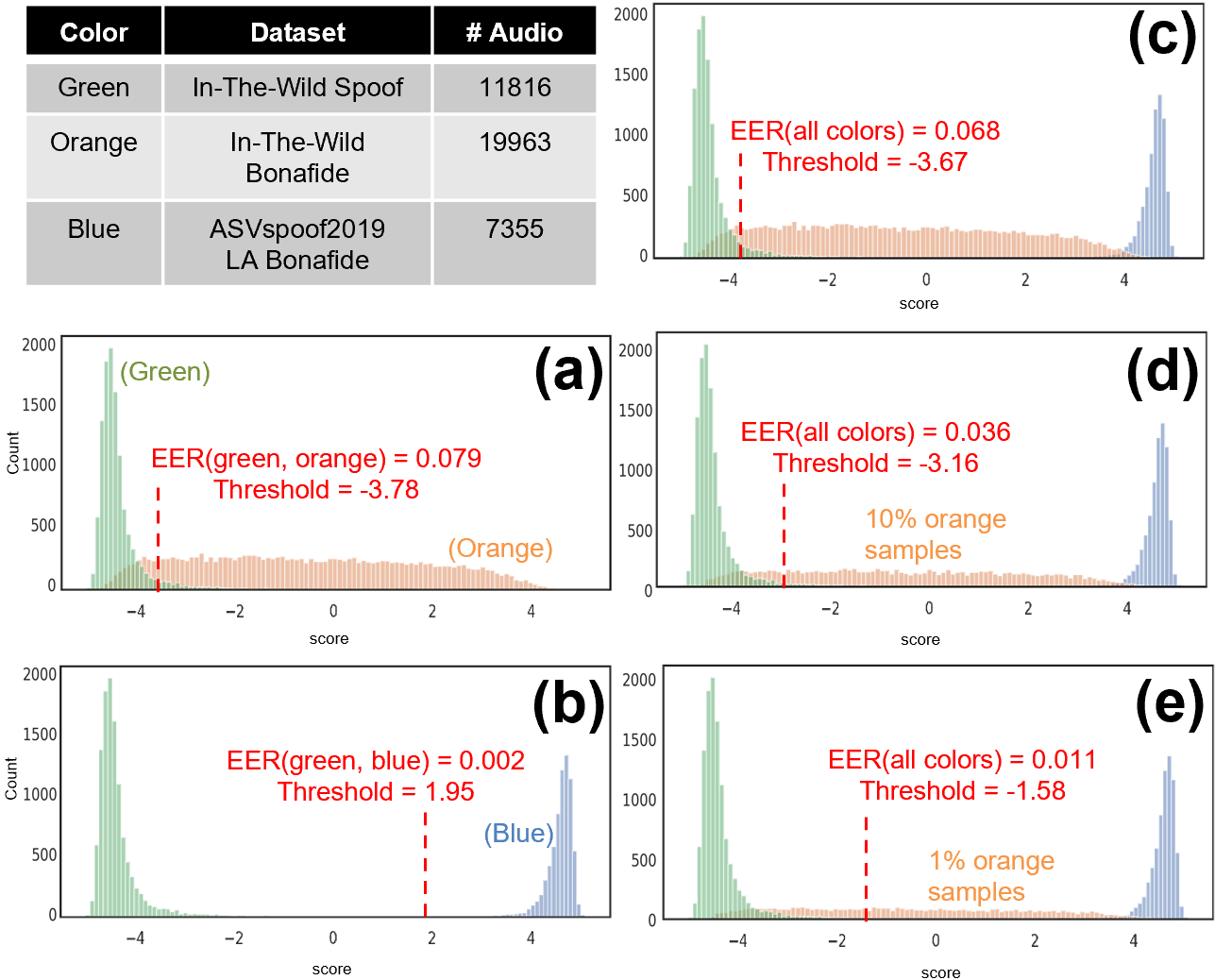}
  \caption{Score distributions from three widely used data subsets. The statistics of the subsets are shown in the top-left table. a) The EER for the green and orange subsets uses threshold $-3.78$. b) If the blue subset is used instead of the orange one, the threshold is $1.95$. c) If the orange and blue subsets are combined, the EER uses threshold $-3.67$. If only d) 10\% or e) 1\% of the orange subset is included in the test set, the threshold is shifted more towards $1.95$. This shows that underrepresented subsets in combined datasets may exert less influence on the EER threshold.
}
  \label{fig:eer_imbalance}
\end{figure}

\textbf{First, the framework fails to reflect real-world diversity.} Most evaluations are limited to a single bona fide speech type \cite{wu2024codecfake,zhao2024emofake,muller2024mlaad}, often focusing on a single environment and speech style (e.g., clean read speech), ignoring the variability present in real-world audio, such as noisy telephony environments, conversational speech, or regional accents. An obvious solution involves merging ADD datasets with diverse bona fide audio datasets. However, this introduces another challenge: as shown in Fig. \ref{fig:eer_imbalance}, underrepresented subsets in combined datasets exert less influence on the overall EER, as the EER threshold balances false positives and false negatives across the entire dataset rather than accounting for individual subsets\cite{cheng2004method}. 

\textbf{Second, the evaluation framework lacks interpretability}. With the proliferation of audio synthesizers that numbers in the thousands\footnote{\url{https://huggingface.co/models?pipeline_tag=text-to-speech}}, the potential number of test datasets is vast. Spoof cross-testing, which reports EERs for every synthesizer dataset, is infeasible and unwieldy, leading to redundancy and reduced clarity. A common workaround is to evaluate on a small subset of recently published or widely used synthesizer datasets \cite{xie2023learning,casado2023conformer,shim2023multi,shim2024beyond,truong2024temporal}. However, this approach introduces biases: models that perform well on recent or popular datasets may underperform on older or underrepresented ones \cite{chettri2021data}, leaving potential vulnerabilities undetected. \textbf{More critically, the framework does not provide insight into subset-level performance}, specifically whether errors arise from some bona fide types being misclassified as spoof (false positives) or from some spoof types going undetected (false negatives). This lack of granularity makes it impossible to accurately diagnose the sources of prediction errors, hindering efforts to improve ADD models. 

To overcome these limitations, we propose bona fide cross-testing, a novel evaluation framework that complements spoof cross-testing. Bona fide cross-testing evaluates ADD models using $K$ diverse bona fide speech types, sourced from various datasets, including those not specifically designed for ADD. Each synthesizer testset is paired with the $K$ bona fide types to generate $K$ distinct test sets. This approach ensures that EERs are calculated separately for each bona fide type, addressing the imbalance issue inherent in combined datasets. The process is repeated for each of the $M$ synthesizer testset, yielding $M \times K$ EERs.

Given the large number of available audio synthesizers ($M>1000$), we propose maximum pooling to summarize EER results, an idea related to Maximum Error Modeling \cite{lingasubramanian2011maximum}. Specifically, we aggregate the EERs accross all synthesizers and only report the $K$-highest EERs that corresponds to the most challenging synthesizer testset. This reflects the realistic assumption that attackers will exploit synthesizers that are the hardest to detect \cite{jang2014survey}. By incorporating both diversity and interpretability, bona fide cross-testing reveals vulnerabilities in ADD models that traditional evaluation methods often overlook.

Our contributions are four-fold:

\begin{enumerate}
\item  We show that combining multiple data subsets into a single test set is suboptimal because underrepresented subsets may have less influence on the EER threshold. Despite this limitation, previous work \cite{casado2023conformer,truong2024temporal,doan2024balance} still merges synthesizer subsets into a single test set (e.g. ASVspoof2021-DF \cite{liu2023asvspoof}), although the subsets vary in size.
\item We reveal that bona fide cross-testing exposes vulnerabilities in ADD models that are often overlooked by conventional evaluation methods such as spoof cross-testing. Notably, bona fide speech types contribute significantly to the challenges faced by ADD systems, whereas traditional approaches primarily focus on the synthesizer type.
\item We propose a novel evaluation framework for ADD that integrates bona fide cross-testing with spoof cross-testing and aggregates EER across spoof tests. Our approach demonstrates greater robustness and interpretability compared to traditional evaluation methods.
\item We provide codes, datasets, and score files required to replicate benchmark results for over 150 synthesizers and 9 bona fide speech types. To the best of our knowledge, this is the first ADD evaluation framework that compares the performance of different bona fide speech types where both the acoustic environment and speech style varies.
\end{enumerate}

\section{Methodoglogy}
\label{sec:point}
This section begins by presenting the widely used spoof cross-testing evaluation framework. We then propose an extension to this framework, incorporating bona fide cross-testing, and provide justifications for the modifications on the original evaluation approach based on the identified limitations.

\subsection{Two types of errors: false positive and false negative}
\label{sec:two_errors}

Following the ISO/IEC standard~\cite{isoiec2016} and Zhang et al. \cite{zhang2023rangebasedequalerrorrate}, we define \textit{spoof} as the positive class and \textit{bona fide} as the negative class. Under this framework, AAD systems are prone to two types of errors: false positives (FP) and false negatives (FN). \footnote{The terms FA and MD used in \cite{Wu2014}, as well as FR and FA in \cite{wang2022chapter}, correspond to FP and FN in this paper, respectively. Although the terminology differs, the definitions remain consistent within the context of the spoofing scenario.}

\begin{itemize}
    \item FP (False Positive): the number of \textit{bona fide} audio samples misclassified as \textit{spoof}.
    \item FN (False Negative): the number of \textit{spoof} audio samples misclassified as \textit{bona fide}.
\end{itemize}

The normalized (proportional) versions of FP and FN are called the false positive rate (FPR) and false negative rate (FNR), and TP (true positive) and TN (true negative) refer to correctly predicted \textit{spoof} and \textit{bona fide} audio samples, respectively.

\subsection{Equal error rate (EER)}

EER is a commonly used metric in binary classification tasks. It is a threshold-free metric that represents the error rate at the specific threshold where the FPR is equal to the FNR (or closest to the FNR in the discrete case). FPR $P_\text{FP}(\tau)$ and FNR $P_\text{FN}(\tau)$ are defined following \cite{zhang2023rangebasedequalerrorrate}. In this setting, a nagative class (bona fide) will have a higher classification score and vice versa:

\begin{equation}
    \label{eq:point_fpr}
    \begin{aligned}
    P_\text{FP}(\tau) &= \frac{1}{| \varLambda_{\mathcal{N}}| } \sum_{\phantom{0}j \in \varLambda_{\mathcal{N}} } \mathds{1}(s_j < \tau), \\
    \end{aligned}
\end{equation}

\begin{equation}
    \label{eq:point_fnr}
    \begin{aligned}
    P_\text{FN}(\tau) &= \frac{1}{| \varLambda_{\mathcal{P}}| } \sum_{\phantom{0}j \in \varLambda_{\mathcal{P}} } \mathds{1}(s_j \geq \tau), \\
    \end{aligned}
\end{equation}

where both $P_\text{FP}(\tau)$ and $P_\text{FN}(\tau)$ are functions of a pre-defined threshold $\tau$. $\varLambda_{\mathcal{N}}$ and $\varLambda_{\mathcal{P}}$ are the bona fide and spoof audio subset, respectively. Then, $|\varLambda_{\mathcal{N}}|$ and $|\varLambda_{\mathcal{P}}|$ denote the total number of bona fide and spoof audio samples in each subset. $\mathds{1}(\cdot)$ denotes the indicator function that outputs 1 when the condition is true and 0 otherwise.

EER is decided by $\hat\tau$ where the value of $P_\text{FP} (\hat\tau)$ is infinitesimally close to $P_\text{FN} (\hat\tau)$. Then, EER can be computed by:
\begin{equation}
\label{eq:eer}
    EER = \frac{P_\text{FP} (\hat\tau)  + P_\text{FN} (\hat\tau) }{2}, \\
\end{equation}
where
\begin{equation}
\label{eq:threshold}
    \hat\tau = \arg \min_{\tau} |P_\text{FP} (\tau) - P_\text{FN} (\tau) |.
\end{equation}

\subsection{Spoof cross-testing}
Currently, ADD models are primarily evaluated on a single dataset that combines multiple synthesizers \cite{liu2023asvspoof,muller2024mlaad} or through spoof cross-testing \cite{wu2024codecfake,Wang2019ASVspoof2A,zhao2024emofake}, a method in which a single bona fide speech type (e.g., clean read speech) is paired with $M$ synthesize datasets (each dataset is only generated by one synthesizer) to generate $M$ test sets. An EER is reported for each test set.

Assume we have one bona fide dataset $\varLambda_{\mathcal{N}}$ and $M$ synthesizer datasets $\{\varLambda^{m}_{\mathcal{P}}\}^{M}_{m=1}$, we define the FNR of the $m$-th spoof subset as:

\begin{equation}
    \begin{aligned}
    P_\text{FN}^{m}(\tau) &= \frac{1}{| \varLambda^{m}_{\mathcal{P}}| } \sum_{\phantom{0}j \in \varLambda^{m}_{\mathcal{P}} } \mathds{1}(s_j \geq \tau), \\
    \end{aligned}
\end{equation}

Then, the EER of the test set that combines $\varLambda_{\mathcal{N}}$ and the $m$-th synthesizer dataset $\varLambda^{m}_{\mathcal{P}}$ is:

\begin{equation}
    \mathrm{EER}_m = \frac{P_\text{FP} (\hat\tau_m)  + P^{m}_\text{FN} (\hat\tau_m) }{2}, \\
\end{equation}
where
\begin{equation}
    \hat\tau_m = \arg \min_{\tau} |P_\text{FP} (\tau) - P^{m}_\text{FN} (\tau) |.
\end{equation}

\subsection{Bona fide cross-testing}

As spoof cross-testing is limited to a single bona fide speech type, it ignores the variability present in real-world bona fide audio, such as noisy telephony environments, conversational speech, or regional accents.

To address this, we propose bona fide cross-testing in addition to spoof cross-testing. In addition, as the bona fide speech types in ADD datasets are scarce, we propose to evaluate ADD models using $K$ diverse bona fide speech types collected from various datasets, including those not specific to ADD. 

Assume we have $K$ bona fide datasets $\{\varLambda^{k}_{\mathcal{N}}\}^{K}_{k=1}$ and $M$ synthesizer datasets $\{\varLambda^{m}_{\mathcal{P}}\}^{M}_{m=1}$, we define the FPR of the $k$-th bona fide dataset as:

\begin{equation}
    \begin{aligned}
    P_\text{FP}^{k}(\tau) &= \frac{1}{| \varLambda^{k}_{\mathcal{P}}| } \sum_{\phantom{0}j \in \varLambda^{k}_{\mathcal{P}} } \mathds{1}(s_j \geq \tau), \\
    \end{aligned}
\end{equation}

Then, the EER of the test set that combines the $k$-th bona fide dataset $\varLambda^{k}_{\mathcal{N}}$ and the $m$-th synthesizer dataset $\varLambda^{m}_{\mathcal{P}}$ is:

\begin{equation}
    \mathrm{EER}_{k,m} = \frac{P^{k}_\text{FP} (\hat\tau_{k,m})  + P^{m}_\text{FN} (\hat\tau_{k,m}) }{2}, \\
\end{equation}
where
\begin{equation}
    \hat\tau_{k,m} = \arg \min_{\tau} |P^{k}_\text{FP} (\tau) - P^{m}_\text{FN} (\tau) |.
\end{equation}

\subsection{Maximum pooling on spoof cross-testing results}

Given the large number of available audio synthesizers ($M>1000$), we propose average and maximum pooling to summarize the EER results. Specifically, we aggregate the EER results across the $M$ synthesizer types and to report only the $K$ highest and average EERs. The $K$ highest EERs, which correspond to the most challenging synthesizers, are reported to reflect the realistic assumption that attackers will exploit synthesizers that are the hardest to detect.

Our max-pooled EER (mEER$_k$) is defined as:
\begin{equation}
    \mathrm{mEER}_k = \max_{m} \mathrm{EER}_{k,m}
\end{equation}

We refrain from further aggregating the mEERs across the $K$ bona fide types, as the performance on individual bona fide types is context-dependent and critical. For instance, if an ADD model is deployed to detect fake news, the mEER for bona fide audio in the news domain is more relevant than those for other domains.

\section{Experiment Setup}
\label{sec:exp}


\begin{table}[t]
\setlength{\tabcolsep}{3pt}
  \caption{Bona fide and spoof test subsets used in our evaluation. The total duration (Dur) in minutes for each subset is reported. SR represents audio sampling rate (in kHz).}
  \label{tab:datasets}
  \centering
    \begin{tabular}{clccccc}
  \toprule
  ID & \multirow{1}{*}{\raisebox{-\heavyrulewidth}{Dataset}} & \multirow{1}{*}{Year} & SR & Dur & \# Audio\\
  \midrule
 \multicolumn{6}{c}{Bona fide test subset \scriptsize{\color{gray}(different environments and speech types)}}\\
 \midrule
   & AMI-Meeting \cite{carletta2005ami}& 2006 \\
  $b_1$ & \quad IHM \scriptsize{\color{gray}(meeting)} & & 16 & 521 & 13K\\
  $b_2$ & \quad SDM \scriptsize{\color{gray}(meeting)} & & 16 & 521 & 13k\\
   & LibriSpeech \cite{panayotov2015librispeech} & 2015 \\
  $b_3$ & \quad test-clean \scriptsize{\color{gray}(storybook)} & & 16 & 324 & 2.6K \\
  $b_4$ & \quad test-other \scriptsize{\color{gray}(storybook)} & & 16 & 320 & 2.9K \\
  $b_5$ & VCTK 0.92 \cite{Yamagishi2019CSTRVC} \scriptsize{\color{gray}(news)} & 2019 & 48 & 46 & 755 \\
  $b_6$ & \scalebox{.8}[1.0]{FakeAVCeleb-v1.2}\cite{Khalid2021FakeAVCelebAN} \color{gray}\scalebox{.5}[0.8]{(interview)} & 2021 & 44.1 & 1K & 10K \\
  $b_7$ & In-The-Wild \cite{Mller2022DoesAD} \color{gray}{\scalebox{.6}[0.8]{(social media)}} & 2022 & 16 & 1.2K & 20K \\
  $b_{8}$ & EmoFake-EN \cite{zhao2024emofake} \scriptsize{\color{gray}(emotion)} & 2022 & 16 & 163 & 3.5K \\
  $b_{9}$ & \scalebox{.5}[1.0]AV-Deefake-1M \cite{Cai2023AVDeepfake1MAL} \color{gray}{\scalebox{.5}[0.8]{(interview)}} & \scalebox{.9}[1.0]2024 & 44.1 & 229 & 1.5K\\
  \midrule
 \multicolumn{6}{c}{Spoof test subset \scriptsize{\color{gray}(different synthesizers)}}\\
 \midrule
  $s_1$ & ASVspoof2019 LA \cite{Wang2019ASVspoof2A} & 2019 & 16 & 3.3K & 64K \\
  & \multicolumn{3}{l}{\quad \scriptsize \color{gray}{(13 synthesizers: $s_{\text{1,1}}$ - $s_{\text{1,13}}$)}}\\
  $s_2$ & ASVspoof2021 DF \cite{liu2023asvspoof} & 2021 & 16 & 26K & 519K \\
  & \multicolumn{3}{l}{\quad \scriptsize \color{gray}{(110 synthesizers: $s_{\text{2,1}}$ - $s_{\text{2,110}}$)}}\\
  $s_3$ & FakeAVCeleb-v1.2 \cite{Cai2023AVDeepfake1MAL} & 2021 & 44.1 & 906 & 11K\\
  & \multicolumn{3}{l}{\quad \scriptsize \color{gray}{(1 synthesizer: $s_{\text{3,1}}$)}}\\
  $s_4$ & EmoFake-EN \cite{zhao2024emofake} & 2022 & 16 & 670 & 14K\\
  & \multicolumn{3}{l}{\quad \scriptsize \color{gray}{(5 synthesizers: $s_{\text{4,1}}$ - $s_{\text{4,5}}$)}}\\
  $s_5$ & AV-Deefake-1M \cite{Cai2023AVDeepfake1MAL} & 2024 & 44.1 & 229 & 1.5K\\
  & \multicolumn{3}{l}{\quad \scriptsize \color{gray}{(4 synthesizers: $s_{5,1}$ - $ s_{5,4}$)}}\\
  $s_6$ & CodecFake \cite{wu2024codecfake} & 2024 & 16/24 & 296 & 4.5K\\
  & \multicolumn{3}{l}{\quad \scriptsize \color{gray}{(6 synthesizers: $s_{6,1}$ - $ s_{6,6}$)}}\\
  $s_{7}$ & MLAAD-v3-EN \cite{muller2024mlaad} & 2024 & 22.05 & 2.5K & 19K\\
  & \multicolumn{3}{l}{\quad \scriptsize \color{gray}{(19 synthesizers: $s_{7,1}$ - $ s_{7,19}$)}}\\
  $s_{8}$ & LlamaPartialSpoof \cite{Luong2024LlamaPartialSpoofAL} & 2024 & 16 & 6.9K & 66K\\
  & \multicolumn{3}{l}{\quad \scriptsize \color{gray}{(6 synthesizers: $s_{8,1}$ - $ s_{8,6}$)}}\\
  \bottomrule
\end{tabular}
\end{table}

To improve the robustness of the evaluation, we collect audios from more than 150 synthesizers and 9 bona fide audio datasets as shown in Table \ref{tab:datasets}. In-The-Wild \cite{Mller2022DoesAD} is not included in the spoof test subsets as it uses unknown number of synthesizers.

Among the nine bona fide datasets, five of them are sourced from non-ADD corpora, originally designed for ASR \cite{kwok2024continual,kwok2024continual_2} and TTS \cite{yuen2023asr} tasks: LibriSpeech test-clean (clean US English read speech) \cite{panayotov2015librispeech}, LibriSpeech test-other (noisy and accented speech) \cite{panayotov2015librispeech}, AMI IHM (meeting speech with headset microphones) \cite{carletta2005ami}, AMI SDM (meeting speech with a single distant microphone) \cite{carletta2005ami}, and VCTK (news-domain speech) \cite{Yamagishi2019CSTRVC}.

As underrepresented subsets in combined datasets may exert less influence on the EER threshold, we further partition the existing ADD datasets such that each data subset is generated by one synthesizer only and EERs are computed separately. To enhance reproducibility, we have released a dataset containing 600 audio samples per synthesizer and bona fide speech type, together with the code to reproduce our bona fide cross-testing evaluation results, available at \url{https://empty.com}.

We evaluate the performance of three recent self-supervised learning (SSL) models, chosen for their robustness across varying audio domains. 1) Wav2Vec-Conformer \cite{casado2023conformer}: Built on the pre-trained XLSR model \cite{conneau2020unsupervisedcrosslingualrepresentationlearning}, a variant of wav2vec 2.0, and trained on the ASVspoof2019 LA dataset. 2) Wav2Vec-TCM \cite{truong2024temporal}: An extension of Wav2Vec-Conformer with improved temporal channel dependency modeling. 3) Wav2Vec-SCL \cite{doan2024balance}: Replaces the Conformer in Wav2Vec-Conformer with linear classifiers and uses Supervised Contrastive Learning (SCL) for learning robust representations. It is trained on ASVspoof2019 LA and additional resynthesized data. The sampling rates are further standardized to match the requirements of each model.

\section{Results and Discussions}
\label{sec:res}

\begin{table*}[ht!]
\centering
\caption{EER (\%) results using our bona fide cross-testing evaluation framework. Given the $M \times K$ EERs obtained from bona fide and spoof cross-testing, we report the $K$ highest and average EERs.}
\begin{tabular}{ccccccccccc}
\toprule
            Method  &   $b_{1}$   &   $b_{2}$   &   $b_{3}$   &   $b_{4}$   &   $b_{5}$   &   $b_{6}$   &   $b_{7}$   &   $b_{8}$   &   $b_{9}$ & avg.\\
\midrule
\multicolumn{1}{l}{\textit{max. EER of $M=164$ synthesizers}}\\
    \multicolumn{1}{l}{\quad Wav2Vec-Conformer \cite{casado2023conformer}} & 0.95 & 0.94 & 
0.85 & 0.86 & 0.40 & 0.98 & 0.93 & 0.91 & 0.98 & 0.87\\
    \multicolumn{1}{l}{\quad Wav2Vec-TCM\cite{truong2024temporal}} & 0.93 & 0.88 & 0.83 & 0.84 & 0.45 & 0.98 & 0.92 & 0.92 & 0.99 & 0.86\\
    \multicolumn{1}{l}{\quad Wav2Vec-SCL\cite{doan2024balance}} & 0.73 & 0.90 & 0.43 & 0.69 & 0.31 & 0.75 & 0.65 & 0.57 & 0.72 & 0.64\\
\midrule
\multicolumn{1}{l}{\textit{avg. EER of $M=164$ synthesizers}}\\
   \multicolumn{1}{l}{\quad Wav2Vec-Conformer\cite{casado2023conformer}} & 0.11 & 0.11 & 
0.05 & 0.08 & 0.01 & 0.20 & 0.12 & 0.07 & 0.22 & 0.11\\
   \multicolumn{1}{l}{\quad Wav2Vec-TCM\cite{truong2024temporal}} & 0.14 & 0.14 & 0.04 & 0.06 & 0.01 & 0.15 & 0.09 & 0.07 & 0.15 & 0.09\\
   \multicolumn{1}{l}{\quad Wav2Vec-SCL\cite{doan2024balance}} & 0.07 & 0.14 & 0.03 & 0.10 & 0.01 & 0.12 & 0.08 & 0.04 & 0.12 & 0.08\\
\bottomrule
\end{tabular}
\label{tab:main_result}
\end{table*}

\begin{figure}[t]
  \centering
  \includegraphics[width=\linewidth]{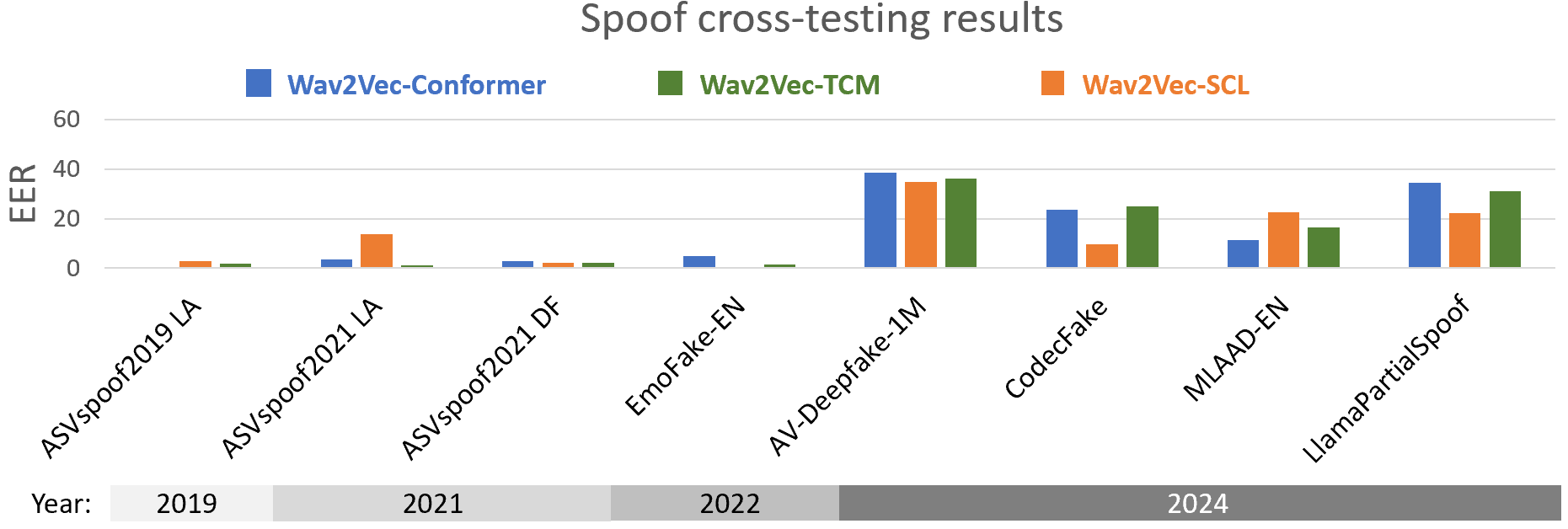}
  \caption{Traditional evaluation results on 8 ADD testsets \cite{Wang2019ASVspoof2A,liu2023asvspoof,Luong2024LlamaPartialSpoofAL,zhao2024emofake,Cai2023AVDeepfake1MAL,wu2024codecfake,muller2024mlaad} by combining multiple synthesizers in a single test set. Limitations: (1) Existing ADD datasets typically contain a single bona fide speech type, limiting real-world diversity. (2) Some test sets, such as ASVspoof2021-DF, have an imbalanced distribution of audio samples across synthesizers. Underrepresented subsets may exert less influence on the EER threshold, leading to unfair evaluation. (3) The evaluation does not provide insight into subset-level performance.
}
  \label{fig:spoof_cross_testing}
\end{figure}

\begin{figure}[t]
  \centering
  \includegraphics[width=0.7\linewidth]{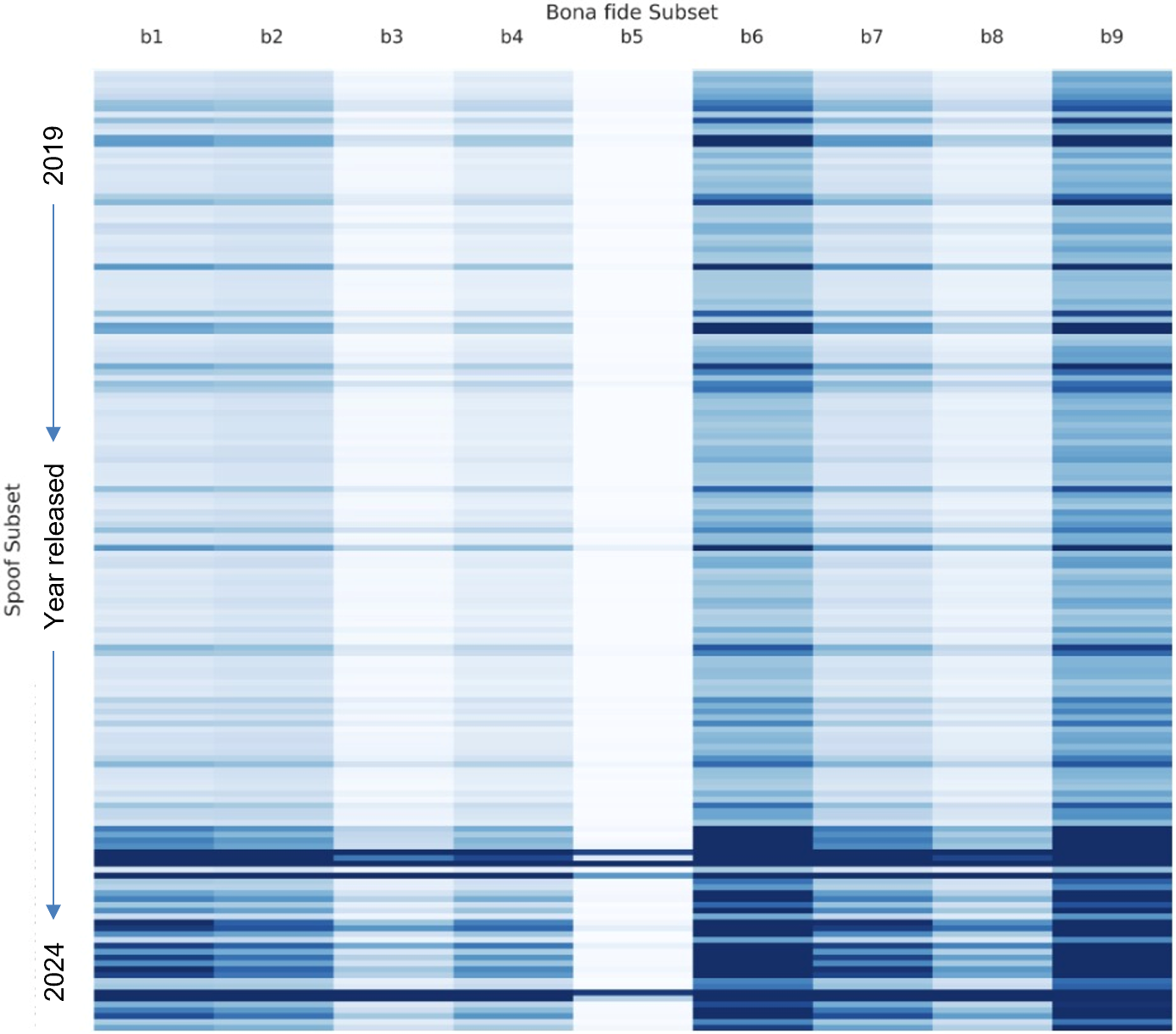}
  \caption{The results of our bona fide cross-testing combined with spoof cross-testing using Wav2Vec-Conformer \cite{casado2023conformer}. The top-left cell is the EER (darker color means higher EER) of the testset that combines spoof subset $s_{1,1}$ with bona fide subset $b_1$ as described in Table \ref{tab:datasets}, and the bottom-right cell combines spoof subset $s_{8,6}$ with bona fide subset $b_9$. The results show that the ADD model performs worse on bona fide subsets b6 and b9 (darker columns), which are celebrity interview speech that may be fast-paced and recorded in a noisy public area.
}
  \label{fig:2d_table}
\end{figure}

We begin by presenting the traditional spoof cross-testing evaluation results on 8 ADD test sets. As shown in Figure \ref{fig:spoof_cross_testing}, all models perform well in datasets created in or before 2022. However, significant performance degradation is observed for most datasets in 2024, indicating that models trained on older datasets struggle to generalize to newer ones.

The analysis is limited as the evaluation approach has two major limitations: (1) It does not capture real-world diversity, as current ADD datasets include only a limited variety of bona fide speech types. (2) It lacks interpretability, failing to reveal whether errors result from bona fide audio being misclassified as spoof (false positives) or spoof audio going undetected (false negatives).

To address this, we present the results of our bona fide cross-testing combined with spoof cross-testing in Figure \ref{fig:2d_table}. The top-left cell is the EER (darker color means higher EER) of the testset that combines spoof subset $s_{1,1}$ with bona fide subset $b_1$ as described in Table \ref{tab:datasets}, and the bottom-right cell combines spoof subset $s_{8,6}$ with bona fide subset $b_9$. The results clearly show that the ADD model performs worse on bona fide subsets $b_6$ and $b_9$, which are celebrity interview speech that may be fast-paced and recorded in a noisy public area. This shows that the poor performance on certain datasets may stem from the difficulty of detecting bona fide audios recorded in complicated environments.

Then, we aggregate the Equal Error Rates (EERs) across the $M=164$ synthesizers and report only the $K=9$ highest and average EERs, as shown in Table \ref{tab:main_result}. The results in the first block indicate that all models fail to detect more than $30\%$ of the spoof audio if only the synthesizer data subset that is the most difficult to synthesizers are considered, The models typically achieve an average EER of approximately 10\%. This shows the importance of reporting the maximum EER, as the results show that although current ADD models generally perform well on most synthesizers, they still contain vulnerabilities towards some of the them.

Finally, the results show that among the three models, Wav2Vec-SCL is the most robust against spoof attacks. Specifically, the first block of Table \ref{tab:main_result} shows that Wav2Vec-SCL generally achieves an EER that is approximately 20\% lower than the other models in the worst-case scenario.

\section{Conclusion}

We introduce bona fide cross-testing, a novel evaluation framework that enhances robustness and interpretability by incorporating diverse bona fide speech types, revealing vulnerabilities overlooked by traditional methods. To support further research, we provide datasets and evaluation codes, aiming to drive the development of more robust and resilient audio deepfake detection models.

\vfill\pagebreak

\section{Acknowledgements}
This research is supported by the National Research Foundation, Singapore and Infocomm Media Development Authority under its Trust Tech Funding Initiative. Any opinions, findings and conclusions or recommendations expressed in this material are those of the author(s) and do not reflect the views of National Research Foundation, Singapore and Infocomm Media Development Authority.

\bibliographystyle{IEEEtran}
\bibliography{mybib}

\begin{thebibliography}{10}
\providecommand{\url}[1]{#1}
\csname url@samestyle\endcsname
\providecommand{\newblock}{\relax}
\providecommand{\bibinfo}[2]{#2}
\providecommand{\BIBentrySTDinterwordspacing}{\spaceskip=0pt\relax}
\providecommand{\BIBentryALTinterwordstretchfactor}{4}
\providecommand{\BIBentryALTinterwordspacing}{\spaceskip=\fontdimen2\font plus
\BIBentryALTinterwordstretchfactor\fontdimen3\font minus \fontdimen4\font\relax}
\providecommand{\BIBforeignlanguage}[2]{{%
\expandafter\ifx\csname l@#1\endcsname\relax
\typeout{** WARNING: IEEEtran.bst: No hyphenation pattern has been}%
\typeout{** loaded for the language `#1'. Using the pattern for}%
\typeout{** the default language instead.}%
\else
\language=\csname l@#1\endcsname
\fi
#2}}
\providecommand{\BIBdecl}{\relax}
\BIBdecl

\bibitem{kwok2025robust}
C.~Y. Kwok, D.-T. Truong, and J.~Q. Yip, ``Robust audio deepfake detection using ensemble confidence calibration,'' in \emph{ICASSP 2025-2025 IEEE International Conference on Acoustics, Speech and Signal Processing (ICASSP)}.\hskip 1em plus 0.5em minus 0.4em\relax IEEE, 2025, pp. 1--5.

\bibitem{liu2023asvspoof}
X.~Liu, X.~Wang, M.~Sahidullah, J.~Patino, H.~Delgado, T.~Kinnunen, M.~Todisco, J.~Yamagishi, N.~Evans, A.~Nautsch \emph{et~al.}, ``Asvspoof 2021: Towards spoofed and deepfake speech detection in the wild,'' \emph{IEEE/ACM Transactions on Audio, Speech, and Language Processing}, vol.~31, pp. 2507--2522, 2023.

\bibitem{muller2024mlaad}
N.~M. M{\"u}ller, P.~Kawa, W.~H. Choong, E.~Casanova, E.~G{\"o}lge, T.~M{\"u}ller, P.~Syga, P.~Sperl, and K.~B{\"o}ttinger, ``Mlaad: The multi-language audio anti-spoofing dataset,'' \emph{arXiv preprint arXiv:2401.09512}, 2024.

\bibitem{wu2024codecfake}
H.~Wu, Y.~Tseng, and H.-y. Lee, ``Codecfake: Enhancing anti-spoofing models against deepfake audios from codec-based speech synthesis systems,'' \emph{arXiv preprint arXiv:2406.07237}, 2024.

\bibitem{Wang2019ASVspoof2A}
\BIBentryALTinterwordspacing
X.~Wang, J.~Yamagishi, M.~Todisco, H.~Delgado, A.~Nautsch, N.~W.~D. Evans, M.~Sahidullah, V.~Vestman, T.~H. Kinnunen, K.~A. LEE, L.~Juvela, P.~Alku, Y.-H. Peng, H.-T. Hwang, Y.~Tsao, H.-M. Wang, S.~L. Maguer, M.~Becker, and Z.~Ling, ``Asvspoof 2019: A large-scale public database of synthesized, converted and replayed speech,'' \emph{Comput. Speech Lang.}, vol.~64, p. 101114, 2019. [Online]. Available: \url{https://api.semanticscholar.org/CorpusID:211532840}
\BIBentrySTDinterwordspacing

\bibitem{zhao2024emofake}
Y.~Zhao, J.~Yi, J.~Tao, C.~Wang, and Y.~Dong, ``Emofake: An initial dataset for emotion fake audio detection,'' in \emph{China National Conference on Chinese Computational Linguistics}.\hskip 1em plus 0.5em minus 0.4em\relax Springer, 2024, pp. 419--433.

\bibitem{cheng2004method}
J.-M. Cheng and H.-C. Wang, ``A method of estimating the equal error rate for automatic speaker verification,'' in \emph{2004 International Symposium on Chinese Spoken Language Processing}.\hskip 1em plus 0.5em minus 0.4em\relax IEEE, 2004, pp. 285--288.

\bibitem{xie2023learning}
Y.~Xie, H.~Cheng, Y.~Wang, and L.~Ye, ``Learning a self-supervised domain-invariant feature representation for generalized audio deepfake detection,'' in \emph{Proc. INTERSPEECH}, vol. 2023, 2023, pp. 2808--2812.

\bibitem{casado2023conformer}
E.~R. Casado, A.~G. Alan{\i}s, A.~G. Garc{\i}a, A.~P. Herreros \emph{et~al.}, ``A conformer-based classifier for variable-length utterance processing in anti-spoofing,'' in \emph{Int. Speech Conf.(INTERSPEECH), Dublin, Ireland}, 2023.

\bibitem{shim2023multi}
H.-j. Shim, J.-w. Jung, and T.~Kinnunen, ``Multi-dataset co-training with sharpness-aware optimization for audio anti-spoofing,'' \emph{arXiv preprint arXiv:2305.19953}, 2023.

\bibitem{shim2024beyond}
H.-j. Shim, M.~Sahidullah, J.-w. Jung, S.~Watanabe, and T.~Kinnunen, ``Beyond silence: Bias analysis through loss and asymmetric approach in audio anti-spoofing,'' \emph{arXiv preprint arXiv:2406.17246}, 2024.

\bibitem{truong2024temporal}
D.-T. Truong, R.~Tao, T.~Nguyen, H.-T. Luong, K.~A. Lee, and E.~S. Chng, ``Temporal-channel modeling in multi-head self-attention for synthetic speech detection,'' \emph{arXiv preprint arXiv:2406.17376}, 2024.

\bibitem{chettri2021data}
B.~Chettri, R.~G. Hautam{\"a}ki, M.~Sahidullah, and T.~Kinnunen, ``Data quality as predictor of voice anti-spoofing generalization,'' \emph{arXiv preprint arXiv:2103.14602}, 2021.

\bibitem{lingasubramanian2011maximum}
K.~Lingasubramanian, S.~M. Alam, and S.~Bhanja, ``Maximum error modeling for fault-tolerant computation using maximum a posteriori (map) hypothesis,'' \emph{Microelectronics Reliability}, vol.~51, no.~2, pp. 485--501, 2011.

\bibitem{jang2014survey}
J.~Jang-Jaccard and S.~Nepal, ``A survey of emerging threats in cybersecurity,'' \emph{Journal of computer and system sciences}, vol.~80, no.~5, pp. 973--993, 2014.

\bibitem{doan2024balance}
T.-P. Doan, L.~Nguyen-Vu, K.~Hong, and S.~Jung, ``Balance, multiple augmentation, and re-synthesis: A triad training strategy for enhanced audio deepfake detection,'' in \emph{Proc. Interspeech 2024}, 2024, pp. 2105--2109.

\bibitem{isoiec2016}
I.~J.~S. Biometrics, ``Iso/iec 30107: Information technology — biometric presentation attack detection,'' 2016.

\bibitem{zhang2023rangebasedequalerrorrate}
\BIBentryALTinterwordspacing
L.~Zhang, X.~Wang, E.~Cooper, N.~Evans, and J.~Yamagishi, ``Range-based equal error rate for spoof localization,'' 2023. [Online]. Available: \url{https://arxiv.org/abs/2305.17739}
\BIBentrySTDinterwordspacing

\bibitem{Wu2014}
Z.~Wu, T.~Kinnunen, N.~Evans, J.~Yamagishi, C.~Hanil{\c{c}}i, M.~Sahidullah, and A.~Sizov, ``{ASVspoof 2015: the First Automatic Speaker Verification Spoofing and Countermeasures Challenge},'' in \emph{Proc. Interspeech}, 2015, pp. 2037--2041.

\bibitem{wang2022chapter}
X.~Wang and J.~Yamagishi, ``A practical guide to logical access voice presentation attack detection,'' in \emph{Frontiers in Fake Media Generation and Detection}.\hskip 1em plus 0.5em minus 0.4em\relax Springer, 2022, pp. 169--214.

\bibitem{carletta2005ami}
J.~Carletta, S.~Ashby, S.~Bourban, M.~Flynn, M.~Guillemot, T.~Hain, J.~Kadlec, V.~Karaiskos, W.~Kraaij, M.~Kronenthal \emph{et~al.}, ``The ami meeting corpus: A pre-announcement,'' in \emph{International workshop on machine learning for multimodal interaction}.\hskip 1em plus 0.5em minus 0.4em\relax Springer, 2005, pp. 28--39.

\bibitem{panayotov2015librispeech}
V.~Panayotov, G.~Chen, D.~Povey, and S.~Khudanpur, ``Librispeech: an asr corpus based on public domain audio books,'' in \emph{2015 IEEE international conference on acoustics, speech and signal processing (ICASSP)}.\hskip 1em plus 0.5em minus 0.4em\relax IEEE, 2015, pp. 5206--5210.

\bibitem{Yamagishi2019CSTRVC}
\BIBentryALTinterwordspacing
J.~Yamagishi, C.~Veaux, and K.~MacDonald, ``Cstr vctk corpus: English multi-speaker corpus for cstr voice cloning toolkit (version 0.92),'' 2019. [Online]. Available: \url{https://api.semanticscholar.org/CorpusID:213060286}
\BIBentrySTDinterwordspacing

\bibitem{Khalid2021FakeAVCelebAN}
\BIBentryALTinterwordspacing
H.~Khalid, S.~Tariq, and S.~S. Woo, ``Fakeavceleb: A novel audio-video multimodal deepfake dataset,'' \emph{ArXiv}, vol. abs/2108.05080, 2021. [Online]. Available: \url{https://api.semanticscholar.org/CorpusID:236976127}
\BIBentrySTDinterwordspacing

\bibitem{Mller2022DoesAD}
\BIBentryALTinterwordspacing
N.~M. M{\"u}ller, P.~Czempin, F.~Dieckmann, A.~Froghyar, and K.~B{\"o}ttinger, ``Does audio deepfake detection generalize?'' \emph{ArXiv}, vol. abs/2203.16263, 2022. [Online]. Available: \url{https://api.semanticscholar.org/CorpusID:247793039}
\BIBentrySTDinterwordspacing

\bibitem{Cai2023AVDeepfake1MAL}
\BIBentryALTinterwordspacing
Z.~Cai, S.~Ghosh, A.~P. Adatia, M.~Hayat, A.~Dhall, and K.~Stefanov, ``Av-deepfake1m: A large-scale llm-driven audio-visual deepfake dataset,'' in \emph{ACM Multimedia}, 2023. [Online]. Available: \url{https://api.semanticscholar.org/CorpusID:265456085}
\BIBentrySTDinterwordspacing

\bibitem{Luong2024LlamaPartialSpoofAL}
\BIBentryALTinterwordspacing
H.-T. Luong, H.~Li, L.~Zhang, K.~A. Lee, and C.~E. Siong, ``Llamapartialspoof: An llm-driven fake speech dataset simulating disinformation generation,'' \emph{ArXiv}, vol. abs/2409.14743, 2024. [Online]. Available: \url{https://api.semanticscholar.org/CorpusID:272827910}
\BIBentrySTDinterwordspacing

\bibitem{kwok2024continual}
C.~Y. Kwok, J.~Q. Yip, and E.~S. Chng, ``Continual learning optimizations for auto-regressive decoder of multilingual asr systems,'' \emph{arXiv preprint arXiv:2407.03645}, 2024.

\bibitem{kwok2024continual_2}
------, ``Continual learning with embedding layer surgery and task-wise beam search using whisper,'' in \emph{2024 IEEE Spoken Language Technology Workshop (SLT)}.\hskip 1em plus 0.5em minus 0.4em\relax IEEE, 2024, pp. 140--146.

\bibitem{yuen2023asr}
C.~Y. Kwok, H.~Y. Li, and E.~S. Chng, ``Asr model adaptation for rare words using synthetic data generated by multiple text-to-speech systems,'' in \emph{2023 Asia Pacific Signal and Information Processing Association Annual Summit and Conference (APSIPA ASC)}.\hskip 1em plus 0.5em minus 0.4em\relax IEEE, 2023, pp. 1771--1778.

\bibitem{conneau2020unsupervisedcrosslingualrepresentationlearning}
\BIBentryALTinterwordspacing
A.~Conneau, A.~Baevski, R.~Collobert, A.~Mohamed, and M.~Auli, ``Unsupervised cross-lingual representation learning for speech recognition,'' 2020. [Online]. Available: \url{https://arxiv.org/abs/2006.13979}
\BIBentrySTDinterwordspacing

\end{thebibliography}

\end{document}